\documentclass[12pt]{article}
\usepackage{amsmath,amssymb,bm,graphicx}
\begin{document}

\begin{flushright}
%April 15, 2011
\end{flushright}

\vskip 0.5 truecm
\begin{center}
{{\Large{\bf  Quantum Discord, CHSH Inequality and\\ Hidden Variables\footnote{Invited talk given at Asia Pacific Conference and Workshop in Quantum Information Science 2012, 3-7 December 2012, Pullman Putrajaya Lakeside, Malaysia
}}}\\
---{\large{\it Critical reassessment of hidden-variables models---}}}
\end{center}

\vskip .5 truecm
\centerline{\bf  Kazuo Fujikawa}
\vskip .4 truecm
\centerline {\it RIKEN Nishina Center, Wako 351-0198, Japan}
\vskip 0.5 truecm

\makeatletter
\@addtoreset{equation}{section}
\def\theequation{\thesection.\arabic{equation}}
\makeatother

\begin{abstract}
Hidden-variables models are critically reassessed. It is first examined if the 
quantum discord is classically described by the hidden-variable model  of Bell in the Hilbert space with $d=2$. The criterion of vanishing quantum discord is related to the notion of reduction and,
surprisingly, the hidden-variable model in $d=2$, which has been believed to be consistent so far, is in fact inconsistent and excluded by the analysis of conditional measurement and reduction. The description of the full contents of quantum discord by the deterministic hidden-variables models is not possible. We also re-examine CHSH inequality. It is shown that the well-known prediction of CHSH inequality $|B|\leq 2$ for the CHSH operator $B$ introduced by Cirel'son is not unique. This non-uniqueness arises from the failure of linearity condition in the non-contextual hidden-variables model in $d=4$ used by Bell and CHSH, in agreement with Gleason's theorem which excludes $d=4$ non-contextual hidden-variables models. If one imposes the linearity condition, their model is converted to a factored product of two $d=2$ models which describes quantum mechanical separable states. The CHSH inequality thus does not test the hidden-variables model in $d=4$. This observation is consistent with an application of the CHSH inequality to quantum cryptography by Ekert, which is based on mixed separable states without referring to dispersion-free representations. As for hidden-variables models, there exist no viable local non-contextual models in any dimensions.
\end{abstract}

\makeatletter
\@addtoreset{equation}{section}
\makeatother
\vspace{2cm}

\section{Introduction}
The notion of quantum discord~\cite{zurek, vedral} is expected to carry information characteristic to quantum mechanics. It is then interesting to examine it in the context of the hidden-variables models such as the ones due to Bell~\cite{bell1} and Kochen-Specker~\cite{kochen} in the Hilbert space with dimensions $d=2$. We may then be able to distinguish the quantum mechanical aspects and classical deterministic aspects of quantum discord. To perform this analysis, it turns out to be  essential to analyze the conditional measurement. Surprisingly, the $d=2$ hidden-variables models such as the ones by Bell and Kochen-Specker, which have been believed to be consistent so far~\cite{beltrametti, peres}, are in fact inconsistent and excluded by this analysis. No consistent non-contextual hidden-variable model exists even in $d=2$ if one requires the uniqueness of the dispersion-free representation of conditional measurement in hidden-variables space, and thus no non-contextual hidden variable models exist in any dimensions~\cite{fujikawa1}. The  criterion of vanishing quantum discord is shown to be reduced to the conditional measurement and reduction, which the conventional hidden-variables models cannot consistently describe, and thus the full contents of quantum discord are not described by the deterministic classical picture~\cite{fujikawa1}, as expected. 

As another issue related to hidden-variables models, it is shown that the well-known prediction of Clauser-Horne-Shimony-Holt (CHSH) inequality $|B|\leq 2$~\cite{bell2, chsh,cs}, for the CHSH operator $B$ introduced by Cirel'son~\cite{cirelson}, is not unique~\cite{fujikawa2}. This non-uniqueness arises from the failure of linearity condition of quantum mechanics in the non-contextual hidden-variables model in $d=4$ used by Bell and CHSH, in agreement with Gleason's theorem which excludes $d=4$ non-contextual hidden-variables models~\cite{gleason}. If one imposes the linearity condition, their model is converted to a factored product of two $d=2$ models which describes quantum mechanical separable states but does not give a non-contextual $d=4$ model. The experimental refutation~\cite{aspect} of CHSH inequality is thus interpreted that the full contents of quantum mechanics even for a far-apart system cannot be described by separable quantum mechanical states only~\cite{fujikawa2}. Also, the non-contextual hidden-variable model in $d=4$ itself which was used to derive the CHSH inequality in the past does not exist. This observation is consistent with an interesting application of the CHSH inequality to quantum cryptography by Ekert~\cite{ekert}, which is based on mixed separable states without referring to dispersion-free representations.

We conclude that there exist no viable local non-contextual hidden-variables models in any dimensions.

\section{ Bell's hidden-variables model in $d=2$}
We start with a brief summary of the hidden-variable model proposed by Bell~\cite{bell1}.
We consider the projection operator
\begin{eqnarray}
P_{{\bf m}}=\frac{1}{2}(1+{\bf m}\cdot{\bf \sigma})
\end{eqnarray}
with a unit vector $|{\bf m}|=1$ and Pauli matrix ${\bf \sigma}$.
The dispersion-free representation of $P_{{\bf m}}$ is defined with the hidden parameter ${\bf \omega}$ in the domain, $\frac{1}{2}\geq \omega\geq -\frac{1}{2}$, by~\cite{beltrametti}
\begin{eqnarray}
P_{{\bf m} \psi}(\omega)=\frac{1}{2}[1+\text{sign}(\omega+\frac{1}{2}|{\bf s}\cdot {\bf m}|)\text{sign}({\bf s}\cdot {\bf m})]
\end{eqnarray}
which assumes the eigenvalues $+1$ or $0$ of the projection operator $P_{{\bf m}}$ for the pure state defined by $|\psi\rangle\langle\psi|=\frac{1}{2}(1+{\bf s}\cdot{\bf \sigma})$ with $|{\bf s}|=1$.

The hidden variable representation $P_{{\bf m} \psi}(\omega)$ reproduces the quantum mechanical result after integration over $\omega$ (with a uniform {\em non-contextual} weight for the hidden variable $\omega$)
\begin{eqnarray}
\int_{-1/2}^{1/2} P_{{\bf m} \psi}(\omega)d\omega =\langle\psi|P_{{\bf m}}|\psi\rangle.
\end{eqnarray}
For a general $2\times 2$ hermitian operator $O$
in a spectral decomposition
$O=\mu_{1}P_{1}+\mu_{2}P_{2}$
with two orthogonal projectors $P_{1}$ and $P_{2}$, $P_{1}+P_{2}=1$, we have the dispersion-free representation
\begin{eqnarray}
O_{\psi}(\omega)=\mu_{1}P_{1,\psi}(\omega)+\mu_{2}P_{2,\psi}(\omega).
\end{eqnarray}
The relations (2.3) and (2.4), which establish the agreement with quantum mechanics in a concrete manner, show that the quantum mechanical linearity condition such as $\langle\psi|(O_{1}+O_{2})|\psi\rangle=\langle\psi|O_{1}|\psi\rangle+\langle\psi|O_{2}|\psi\rangle$ is satisfied after the integration over the hidden variable $\omega$.

\section{ Conditional measurement }
We next analyze the conditional measurement in the framework of hidden-variables models.
Starting with the initial state $\rho$, suppose that we first measure the projector $B$ and then measure the projector $A$ with $AB\neq0$. We define
\begin{eqnarray}
\rho_{B}\equiv \frac{B\rho B}{\text{Tr}\rho B}, \hspace{1cm} \text{Tr}\rho B \neq 0,
\end{eqnarray}
and then the probability of the measurement of $A$ is given by 
\begin{eqnarray}
\text{Tr}[\rho_{B}A]=\frac{\text{Tr}[(B\rho B)A]}{\text{Tr}[\rho B]}.
\end{eqnarray}
This construction is faithful to the original quantum mechanical definition of the conditional measurement.

The projected state $\rho_{B}$ corresponds to $|\psi_{B}\rangle\langle\psi_{B}|=B$ in a matrix notation, and in Bell's construction we have the dispersion free representation (with $A=P_{\vec{m}}$, $B=P_{\vec{n}}$)
\begin{eqnarray}
A_{\psi_{B}}(\omega)=\frac{1}{2}[1+\text{sign}(\omega+\frac{1}{2}|\vec{n}\cdot\vec{m}|)\text{sign}(\vec{n}\cdot\vec{m})]
\end{eqnarray}
which is symmetric in $A$ and $B$, and we obtain the identical expression for $B_{\psi_{A}}(\omega)$.

An alternative way  is to {\em define} the ratio of averages~\cite{umegaki, davies}
\begin{eqnarray}
\alpha_{B}(A)=\frac{\text{Tr}\rho (BAB)}{\text{Tr}[\rho B]}, \hspace{1cm} \text{Tr}[\rho B]\neq 0
\end{eqnarray}
as the conditional probability measure of $A$ after the measurement of $B$.
Here we emphasize a new composite {\em operator} $BAB$, which is no more a projection operator, while we emphasized the modification of the {\em state} in (3.1) before. These two are naturally identical in quantum mechanics.

For the projectors, we have
\begin{eqnarray}
P_{\vec{n}}P_{\vec{m}}P_{\vec{n}}=\frac{1}{2}(1+\vec{n}\cdot\vec{m})
P_{\vec{n}}
\end{eqnarray}
and $P_{\vec{m}}P_{\vec{n}}P_{\vec{m}}=\frac{1}{2}(1+\vec{n}\cdot\vec{m})P_{\vec{m}}$. 
We then obtain the dispersion free representation, 
\begin{eqnarray}
\frac{(BAB)_{\psi}(\omega)}{\langle \psi|B|\psi\rangle}&=&\frac{(1+\vec{n}\cdot\vec{m})}{(1+\vec{n}\cdot\vec{s})}\nonumber\\
&\times&\frac{1}{2}[1+\text{sign}(\omega+\frac{1}{2}|\vec{s}\cdot\vec{n}|)\text{sign}(\vec{s}\cdot\vec{n})]
\end{eqnarray}
using  $B_{\psi}(\omega)$ with  $A=P_{\vec{m}}$ and $B=P_{\vec{n}}$. 

One then confirms that the conditional measurement is consistently described by either way, in agreement with the quantum mechanical result as
\begin{eqnarray}
\frac{\text{Tr}[\rho BAB]}{\text{Tr}[\rho B]}
&=&\int d\omega A_{\psi_{B}}(\omega)\nonumber\\
&=&\int d\omega\frac{(BAB)_{\psi}(\omega)}
{\langle \psi|B|\psi\rangle}\nonumber\\
&=&\frac{(1+\vec{n}\cdot\vec{m})}{2},
\end{eqnarray}
which also agrees with $\text{Tr}[\rho ABA]/\text{Tr}[\rho A]$.

In passing, this example shows that the conditional measurement in hidden-variables models does not follow the classical conditional probability rule  for general non-commuting $A$ and $B$,
\begin{eqnarray}
\frac{\text{Tr}[\rho BAB]}{\text{Tr}[\rho B]}
 \neq \frac{\mu[a_{\rho}\cap b_{\rho}]}{\mu[ b_{\rho}]}.
\end{eqnarray}
Here $\mu[ a_{\rho}]$ stands for the probability of $A$ for the state $\rho$ 
in a generic notation 
\begin{eqnarray}
\mu[a_{\rho}]\equiv \int_{\Omega}A_{\rho}(\omega)d\mu(\omega) =\text{Tr}[\rho A],
\end{eqnarray}
where
\begin{eqnarray}
a_{\rho}=A_{\rho}^{-1}(1)=\{\omega\in \Omega : A_{\rho}(\omega)=1\}.
\end{eqnarray}
The classical conditional probability rule which satisfies $\mu[a_{\rho}\cap b_{\rho}]=\mu[b_{\rho}\cap a_{\rho}]$, if imposed on noncontextual hidden-variables models, eliminates the crucial notion of {\em reduction in quantum mechanics}, as is seen by the fact that $a_{\rho}$ and $b_{\rho}$ in $\mu[a_{\rho}\cap b_{\rho}]$ are defined by the same original state $\rho$ although $\mu[a_{\rho}\cap b_{\rho}]$ is divided by $\mu[ b_{\rho}]$ (Bayes rule). 
\\

We recognize that the two dispersion-free representations
\begin{eqnarray}
A_{\psi_{B}}(\omega)=\frac{1}{2}[1+\text{sign}(\omega+\frac{1}{2}|\vec{n}\cdot\vec{m}|)\text{sign}(\vec{n}\cdot\vec{m})],
\end{eqnarray}
and
\begin{eqnarray}
\frac{(BAB)_{\psi}(\omega)}{\langle \psi|B|\psi\rangle}&=&\frac{(1+\vec{n}\cdot\vec{m})}{(1+\vec{n}\cdot\vec{s})}\nonumber\\
&\times&\frac{1}{2}[1+\text{sign}(\omega+\frac{1}{2}|\vec{s}\cdot\vec{n}|)\text{sign}(\vec{s}\cdot\vec{n})],
\end{eqnarray}
lead to two conflicting dispersion-free representations in the  hidden variables space parameterized by $\omega$ for the {\em identical} quantum mechanical object $\text{Tr}[\rho BAB]/\text{Tr}[\rho B]$, although both of them reproduce the same quantum mechanical result after averaging over the hidden variable. 

We now postulate that {\em any physical quantity should have a unique expression in the hidden variables space}, just as any quantum mechanical quantity has a unique space-time dependence.  
This requirement is not satisfied by the expression of the 
conditional measurement in the $d=2$ non-contextual hidden variables model of Bell, which in turn implies that the {\em reduction} and {\em state preparation} are not consistently described in the hidden-variables model. (It is shown that the same conclusion holds for the model of Kochen-Specker~\cite{kochen} also.)
After all, one of the main purposes of the hidden-variables model is to avoid the sudden state reduction.

The representation (3.11) based on (3.2) incorporates the reduction and state preparation, but the reduction is not explicit in the alternative construction (3.12). To be precise, the representation (3.11), although it satisfies the condition of reduction in a quantum mechanical sense, does not specify the reduction in the sense of dispersion-free representation since the hidden parameter $\omega$ after the first measurement of $B$ is not specified in the construction. To resolve this issue, one needs to go beyond the conventional construction of the hidden-variables representation by allowing, for example, the opening of a new hidden-variables space every time one makes a new measurement~\cite{fujikawa1, fujikawa3} such as
\begin{eqnarray}
A_{\psi_{B}}(\omega^{\prime})B_{\psi}(\omega)/\int d\omega B_{\psi}(\omega)
&=&\frac{1}{2}[1+\text{sign}(\omega^{\prime}+\frac{1}{2}|\vec{n}\cdot\vec{m}|)\text{sign}(\vec{n}\cdot\vec{m})]
\\
&\times&\frac{1}{2}[1+\text{sign}(\omega+\frac{1}{2}|\vec{s}\cdot\vec{n}|)\text{sign}(\vec{s}\cdot\vec{n})]\frac{2}{(1+\vec{n}\cdot\vec{s})},\nonumber
\end{eqnarray}
namely, the first measurement of $B$ introduces $\omega$ and the second measurement of $A$ introduces $\omega^{\prime}$. One then
later integrates over $\omega$ and $\omega^{\prime}$ {\em independently}.
By this way one can incorporate the notion which corresponds to reduction into the dispersion-free representation, but the construction goes beyond the conventional idea of hidden-variables models.\\

From a point of view of the dual structure of operator and state $(O,\rho)$ in quantum mechanics, we have two equivalent choices
\begin{eqnarray}
(A,B\rho B)\ \ {\rm or}\ \ (BAB,\rho),
\end{eqnarray}
respectively, before moving to hidden variables models. These two are obviously equivalent in quantum mechanics (or in any trace representation with density matrix), but they are quite different in Bell's construction due to  the lack of definite {\em associative properties} of various operations.
An interesting example is given by the measurement of $A$ immediately after the measurement of $A$:
One prescription (3.3) gives an $\omega$ independent unit representation, while the other (3.6) gives $A_{\psi}(\omega)/\int A_{\psi}(\omega)d\omega$ which has the same $\omega$ dependence as the first measurement of $A$.

We may thus conclude that the {\em hidden-variables model cannot describe the conditional measurement consistently}~\cite{fujikawa1}.

\section{Quantum discord for a two-partite system}

We now examine if the quantum discord (a measure of the quantum excess of correlations) is described by the hidden-variables model in Sections 2 and 3. If this description is possible, it would imply that a deterministic classical description of quantum discord is possible.
The notion of quantum discord for a two-partite system described by the density matrix $\rho_{XY}$ is defined as a difference of the quantum conditional entropy~\cite{zurek, vedral}
\begin{eqnarray}
\sum_{j}p_{j}S(\rho_{Y|\Pi^{X}_{j}})
\end{eqnarray}
and the formal conditional entropy $S(X,Y)-S(X)$ as, 
\begin{eqnarray}
D\equiv\sum_{j}p_{j}S(\rho_{Y|\Pi^{X}_{j}})
-[S(X,Y)-S(X)],
\end{eqnarray}
where the general definition of entropy is $S(\rho)=-{\rm Tr}\rho\ln\rho$ for a given state $\rho$, and 
$\rho_{X}={\rm Tr}_{Y}\rho_{XY}$. The important property is that $\sum_{j}p_{j}S(\rho_{Y|\Pi^{X}_{j}})$ involves the measurement process while $S(X,Y)-S(X)$ does not contain any measurement process.
Actually, the quantum discord $D$ is defined at the minimum of the first term in (4.2) with respect to all the possible choices of the set of projectors $\{\Pi^{X}_{j}\}$. 
The orthogonal projectors are defined by
\begin{eqnarray}
\Pi^{X}_{i}\Pi^{X}_{j}=\Pi^{X}_{j}\Pi^{X}_{i}=\delta_{i,j}\Pi^{X}_{j},\ \ \sum_{j}\Pi^{X}_{j}=1,
\end{eqnarray}
and  the normalized density matrix after the measurement of $\{\Pi^{X}_{j}\}$,
\begin{eqnarray}
\rho_{Y|\Pi^{X}_{j}}=\frac{\text{Tr}_{X}[(\Pi^{X}_{j}\otimes 1)\rho_{XY}(\Pi^{X}_{j}\otimes 1)]}{p_{j}}
\end{eqnarray}
with 
\begin{eqnarray}
p_{j}=\text{Tr}[(\Pi^{X}_{j}\otimes 1)\rho_{XY}]. 
\end{eqnarray}

At this point it is instructive to consider a classical system defined by the density matrix
\begin{eqnarray}
\rho=\rho(x,y)
\end{eqnarray}
where $x$ and $y$ collectively stand for the phase space variables of X-system and Y-system, respectively. Then $\rho(x)=\int\rho(x,y)dy$ and one may identify
\begin{eqnarray}
&&p_{j}\leftrightarrow \Delta x_{j}\rho(x_{j})=\int\Delta x_{j}\rho(x_{j},y)dy, \nonumber\\
&&\rho_{Y|\Pi^{X}_{j}}\leftrightarrow \frac{\rho(x_{j},y)}{\rho(x_{j})},
\end{eqnarray}
where $\Delta x_{j}$ specifies a specific domain in the phase space. In this identification, we have
\begin{eqnarray}
S(X,Y)-S(X)&\rightarrow& -\int dxdy\rho(x,y)\ln\rho(x,y)+\int dx\rho(x)\ln\rho(x)\nonumber\\
&=&-\int dxdy\rho(x,y)\ln\left(\rho(x,y)/\rho(x)\right)
\end{eqnarray}
while
\begin{eqnarray}
\sum_{j}p_{j}S(\rho_{Y|\Pi^{X}_{j}})&\rightarrow& -\sum_{j}\int dy\Delta x_{j}\rho(x_{j},y)\ln\frac{\rho(x_{j},y)}{\rho(x_{j})}\nonumber\\
&=&-\int dxdy\rho(x,y)\ln\frac{\rho(x,y)}{\rho(x)},
\end{eqnarray}
and thus $D=0$. Namely, the non-vanishing quantum discord $D$ implies that a certain quantum property is involved.

The crucial property of the quantum discord $D$ thus defined is that it survives even for the separable system without entanglement~\cite{zurek, vedral}.

The following general properties of the quantum discord are known
\begin{eqnarray}
D=0 \Longleftrightarrow \rho_{XY}&=&\sum_{j}\Pi^{X}_{j}\rho_{XY}\Pi^{X}_{j}
\nonumber\\
&=&\sum_{j}p_{j}\Pi^{X}_{j}\otimes\rho^{Y}_{j},
\end{eqnarray}
for a suitable set of projectors $\{\Pi^{X}_{j}\}$, and 
\begin{eqnarray}
0\leq D\leq S(\rho_{X}).
\end{eqnarray}
The proofs of those properties are involved and the readers are referred to original references~\cite{zurek, datta1, dakic, datta2}. 

Intuitively, the apparatus of the quantum measurement, which executes a precise projective measurement of an operator $B(X)$ for the X-system with spectral decomposition
\begin{eqnarray}
B(X)=\sum_{j}b_{j}\Pi^{X}_{j},
\end{eqnarray}
induces the change of the state $\rho_{XY}$ to $\sum_{j}\Pi^{X}_{j}\rho_{XY}\Pi^{X}_{j}$ after the measurement~\cite{neumann}. Thus the first relation in (4.10) shows that this measurement does not change the initial (in general, mixed) state. A characteristic property of quantum mechanics is that the measurement changes the state, for example, a pure state $\rho$ is transformed to a mixed state $\sum_{j}\Pi^{X}_{j}\rho\Pi^{X}_{j}$. Also the final state depends on which kind of  projective measurement one performs, and this is the reason why we look for the minimum of the first term in (4.2).

\subsection{Vanishing condition of the quantum discord}
To establish the vanishing quantum discord specified by (4.10), it is necessary to show 
\begin{eqnarray}
\text{Tr}_{X}A_{X}\rho_{XY}=\sum_{j}\text{Tr}_{X}A_{X}\Pi^{X}_{j}\rho_{XY}\Pi^{X}_{j},
\end{eqnarray}
for {\em any} projector $A_{X}$. It is obvious that the condition (4.10) implies the relation (4.13). Conversely, one may choose any projection operator of the form $A_{X}=|\psi_{X}\rangle\langle\psi_{X}|$, then (4.13) implies
\begin{eqnarray}
\langle\psi_{X}|\rho_{XY}|\psi_{X}\rangle=\langle\psi_{X}|\sum_{j}\Pi^{X}_{j}\rho_{XY}\Pi^{X}_{j}|\psi_{X}\rangle
\end{eqnarray}
for any $|\psi_{X}\rangle$ which in turn implies (4.10); the equality of the average  for {\em any} state $|\psi_{X}\rangle$ in fact implies the equality of the operators themselves.
We thus have to deal with general positive operators 
, $\Pi^{X}_{j}A_{X}\Pi^{X}_{j}$, to discuss the criterion of the vanishing quantum discord.

 In the case of a separable mixed state in $d=4=2\times2$,
\begin{eqnarray}
\rho_{XY}=\sum_{k}w_{k}\rho^{(k)}_{X}\otimes\rho^{(k)}_{Y},
\end{eqnarray}
we have the vanishing condition of the quantum discord
\begin{eqnarray}
\sum_{k}w_{k}[\text{Tr}_{X}A_{X}\rho^{(k)}_{X}]\rho^{(k)}_{Y}
=\sum_{k}w_{k}\sum_{j}[\text{Tr}_{X}A_{X}\Pi^{X}_{j}\rho^{(k)}_{X}\Pi^{X}_{j}]\rho^{(k)}_{Y},
\end{eqnarray}
for any projector $A_{X}$ and a suitable set of projectors $\{\Pi^{X}_{j} \}$.
It is then interesting to examine this condition in Bell's hidden-variables model. The quantity 
\begin{eqnarray}
[\text{Tr}_{X}A_{X}\rho^{(k)}_{X}]
\end{eqnarray}
on the left-hand side has a well-defined 
hidden-variables representation of Bell for $d=2$ in Section 2. 
But two different hidden-variables prescriptions for the quantity on the right-hand side 
\begin{eqnarray}
\text{Tr}_{X}[A_{X}\Pi^{X}_{j}\rho^{(k)}_{X}\Pi^{X}_{j}]
\end{eqnarray}
lead to two different conflicting representations in the hidden-variables space, as explained in Section 3.
Quantum mechanically equivalent expressions lead to quite different dispersion-free expressions in the hidden-variables space in Bell's construction, and there appears to be no clear physical criterion to resolve the ambiguity.
 One may thus conclude that the description  of the criterion of vanishing quantum discord in the {\em hidden-variables space} is ill-defined in  Bell's construction. In this sense, the deterministic description of the full contents of quantum discord is not possible.
\\
  
On the basis of Bell's explicit construction in $d=2$~\cite{bell1}, it was pointed out
that the description of the criterion of vanishing quantum discord in the hidden-variables space is ill-defined. The same conclusion applies to the $d=2$ model by Kochen and Specker~\cite{kochen}. Also, the criterion of the vanishing quantum discord itself (4.10) is related to the notion of the reduction associated with precise projective measurements. The hidden-variables models have difficulties in the description of reduction and state preparation, since the main motivation of hidden-variables models is to avoid the sudden reduction of states, to begin with. 

As for the possible practical applications of quantum discord, other aspects of quantum discord also play an important role. For example, the pure separable state $\rho_{XY}=\rho_{X}\otimes\rho_{Y}$ after the diagonalization of the first factor $\rho_{X}$ satisfies the criterion of vanishing quantum discord (4.10). The non-vanishing quantum discord of the separable mixed state $\rho_{XY}=\sum_{k}w_{k}\rho^{(k)}_{X}\rho^{(k)}_{Y}$  arises from the fact that one cannot diagonalize all $\rho^{(k)}_{X}$ simultaneously. Thus one can store certain "quantum information" in the set $\{\rho^{(k)}_{X}\}$ which are labeled by the associated $\{\rho^{(k)}_{Y}\}$. A salient feature here is that we have no quantum coherence or interference among different members of $\{\rho^{(k)}_{X}\}$, nevertheless this is a quantum mechanical effect.  

We may thus conclude that {\em one of the essential aspects of the  "quantumness" of quantum discord is traced to the reduction of states, i.e., measurement changes physical states}, in contrast to the locality in the analysis of entanglement.

\section{ CHSH inequality: Criterion of entanglement}

For four independent  {\em dichotomic} variables $\{a_{j}, \  a^{\prime}_{j},\ b_{j}, \  b^{\prime}_{j} \}$
 which assume $\pm 1$, one can confirm the relation
\begin{eqnarray}
a_{j}(b_{j}+b^{\prime}_{j})+a^{\prime}_{j}(b_{j}-b^{\prime}_{j})=\pm 2.
\end{eqnarray}
We imagine a two-partite system, and the variables $\{a_{j}, \  a^{\prime}_{j}\}$
belong to the a-system and $\{b_{j}, \  b^{\prime}_{j}\}$ belong to the b-system, respectively. Summing  with any {\em uniform} weight factor $P_{j}\geq 0$ which satisfies  $\sum_{j}P_{j}=1$, we obtain the CHSH inequality~\cite{peres}
\begin{eqnarray}
|\langle ab\rangle+\langle ab^{\prime}\rangle+\langle a^{\prime}b\rangle-\langle a^{\prime}b^{\prime}\rangle|\leq 2
\end{eqnarray}
where $\langle ab\rangle=\sum_{j}P_{j}a_{j}b_{j}$.
The uniform weight for all the combinations of dichotomic variables manifests the strict locality which also implies  non-contextuality.  Entanglement implies that the combination $a_{j}b_{j}$ is more favored than 
$a^{\prime}_{j}b_{j}$, for example, but the relation (5.2) completely ignores this "contextuality". 

Quantum CHSH operator introduced by Cirel'son reads~\cite{cirelson}
\begin{eqnarray}
B={\bf a}\cdot {\bf \sigma}\otimes ({\bf b}+{\bf b}^{\prime})\cdot {\bf \sigma} +{\bf a}^{\prime}\cdot{\bf \sigma}\otimes ({\bf b}-{\bf b}^{\prime})\cdot{\bf \sigma}
\end{eqnarray}
with unit 3-dimensional vectors ${\bf a},{\bf a}^{\prime},{\bf b},{\bf b}^{\prime}$ and the Pauli matrices ${\bf \sigma}$ for a system of two spin-$1/2$ particles, regarded as a $d=4$ dimensional system in the Hilbert space. One can then confirm
\begin{eqnarray}
||B||\leq 2\sqrt{2}
\end{eqnarray}
by noting 
\begin{eqnarray}
&&||{\bf a}\cdot {\bf \sigma}\otimes ({\bf b}+{\bf b}^{\prime})\cdot {\bf \sigma}||\leq |{\bf b}+{\bf b}^{\prime}|,\nonumber\\
&&||{\bf a}^{\prime}\cdot{\bf \sigma}\otimes ({\bf b}-{\bf b}^{\prime})\cdot{\bf \sigma}||\leq |{\bf b}-{\bf b}^{\prime}|,
\end{eqnarray}
and $2 \leq |{\bf b}+{\bf b}^{\prime}|+|{\bf b}-{\bf b}^{\prime}|\leq 2\sqrt{2}$.

For any separable pure state $|\psi\rangle$ such as $|\psi\rangle=|\frac{1}{2}\rangle\otimes|\frac{1}{2}\rangle$, one has in the notation of (5.1)
\begin{eqnarray}
\langle \psi|B|\psi\rangle=a_{z}(b_{z}+b^{\prime}_{z})+a^{\prime}_{z}(b_{z}-b^{\prime}_{z})
\end{eqnarray}
but all the variables are now limited in the domain $[-1, 1]$. Since 
$\langle \psi|B|\psi\rangle$ is a linear function of $a_{z}$ and  $a^{\prime}_{z}$, its absolute value assumes the maximum at the boundary of the domain. By examining $\langle \psi|B|\psi\rangle$ at the four corners of the space $\{a_{z}, a^{\prime}_{z} \}$ one can confirm 
\begin{eqnarray}
|\langle \psi|B|\psi\rangle|\leq 2.
\end{eqnarray}

\subsection{Local non-contextual hidden-variables model of Bell and CHSH}
For any pure $4\times 4$ state
$\rho=|\psi\rangle\langle \psi|$, the local non-contexual hidden-variables model of Bell and CHSH is defined by~\cite{bell2, chsh, cs}
\begin{eqnarray}
\langle {\bf a}\cdot {\bf \sigma}\otimes {\bf b}\cdot {\bf \sigma}\rangle_{\psi}
=\int_{\Lambda} P(\lambda)a_{\psi}(\theta,\lambda)b_{\psi}(\varphi,\lambda)d\lambda
\end{eqnarray}
with dichotomic variables $a_{\psi}(\theta,\lambda)$ and $b_{\psi}(\varphi,\lambda)$. See eq.(2) in the original paper of Bell~\cite{bell2} and eq.(3.5) in the review by Clauser and Shimony~\cite{cs}. See also Werner~\cite{werner}. The parameters $\lambda$ collectively stand for hidden-variables, and  $\theta$ and $\varphi$ stand for the azimuthal angles in the planes perpendicular to the line connecting the two spin systems.

We first analyze the CHSH operator by re-writing it for {\em non-collinear} ${\bf b}$ and ${\bf b}^{\prime}$ as 
\begin{eqnarray}
B&=&{\bf a}\cdot {\bf \sigma}\otimes ({\bf b}+{\bf b}^{\prime})\cdot {\bf \sigma} +{\bf a}^{\prime}\cdot{\bf \sigma}\otimes ({\bf b}-{\bf b}^{\prime})\cdot{\bf \sigma}\nonumber\\
&=& |{\bf b}+{\bf b}^{\prime}|[{\bf a}\cdot {\bf \sigma}\otimes \tilde{{\bf b}}\cdot {\bf \sigma}] +
|{\bf b}-{\bf b}^{\prime}|[{\bf a}^{\prime}\cdot{\bf \sigma}\otimes \tilde{{\bf b}}^{\prime}\cdot{\bf \sigma}]
\end{eqnarray}
by defining two unit vectors
\begin{eqnarray}
\tilde{{\bf b}}=\frac{{\bf b}+{\bf b}^{\prime}}{|{\bf b}+{\bf b}^{\prime}|}, \ \  
\tilde{{\bf b}}^{\prime}=\frac{{\bf b}-{\bf b}^{\prime}}{|{\bf b}-{\bf b}^{\prime}|}, \ \ \tilde{{\bf b}}\cdot\tilde{{\bf b}}^{\prime}=0.   
\end{eqnarray}
By using the above non-contextual hidden-variables formula, we obtain
\begin{eqnarray}
\langle B \rangle_{\psi}
&=&\int P(\lambda)d\lambda [|{\bf b}+{\bf b}^{\prime}|a_{\psi}(\theta,\lambda)\tilde{b}_{\psi}(\phi,\lambda)\nonumber\\
&&+
|{\bf b}-{\bf b}^{\prime}|a_{\psi}(\theta^{\prime},\lambda)\tilde{b}^{\prime}_{\psi}(\phi^{\prime},\lambda)].
\end{eqnarray}
By noting 
\begin{eqnarray}
&& |[|{\bf b}+{\bf b}^{\prime}|a_{\psi}(\theta,\lambda)\tilde{b}_{\psi}(\phi,\lambda)
+
|{\bf b}-{\bf b}^{\prime}|a_{\psi}(\theta^{\prime},\lambda)\tilde{b}^{\prime}_{\psi}(\phi^{\prime},\lambda)]|\nonumber\\
&&\leq [ |{\bf b}+{\bf b}^{\prime}|+|{\bf b}-{\bf b}^{\prime}|]
\end{eqnarray}
and $2 < |{\bf b}+{\bf b}^{\prime}|+|{\bf b}-{\bf b}^{\prime}|\leq 2\sqrt{2}$
 for non-collinear ${\bf b}$ and ${\bf b}^{\prime}$,
we conclude
\begin{eqnarray}
|\langle B \rangle_{\psi}|\leq 2\sqrt{2}.
\end{eqnarray}
To achieve the above upper bound, some domain in hidden-variables space with 
\begin{eqnarray}
a_{\psi}(\theta,\lambda)\tilde{b}_{\psi}(\phi,\lambda)= a_{\psi}(\theta^{\prime},\lambda)\tilde{b}^{\prime}_{\psi}(\phi^{\prime},\lambda)=1
\end{eqnarray}
or 
\begin{eqnarray}
a_{\psi}(\theta,\lambda)\tilde{b}_{\psi}(\phi,\lambda)=a_{\psi}(\theta^{\prime},\lambda)\tilde{b}^{\prime}_{\psi}(\phi^{\prime},\lambda)=-1
\end{eqnarray}
is essential. 
If one assumes otherwise; namely,
if $ a_{\psi}(\theta,\lambda)\tilde{b}_{\psi}(\phi,\lambda)=\pm 1 $ should always imply $a_{\psi}(\theta^{\prime},\lambda)\tilde{b}^{\prime}_{\psi}(\phi^{\prime},\lambda)=\mp1$ for any $\lambda$, respectively, the hidden-variables formula  would always imply for a sum of two {\em non-commuting} operators 
\begin{eqnarray}
\langle {\bf a}\cdot {\bf \sigma}\otimes \tilde{{\bf b}}\cdot {\bf \sigma}\rangle_{\psi}
 +\langle {\bf a}^{\prime}\cdot{\bf \sigma}\otimes \tilde{{\bf b}}^{\prime}\cdot{\bf \sigma}\rangle_{\psi}=0.
\end{eqnarray}
This does not hold for  generic quantum states $\psi$, and thus the above upper bound is generally achieved if the hidden-variables model should make sense.

\subsection{ Conventional CHSH inequality} 
The conventional CHSH inequality is based on the evaluation~\cite{chsh}
\begin{eqnarray}
\langle B\rangle_{\psi}&=& \langle {\bf a}\cdot {\bf \sigma}\otimes ({\bf b}+{\bf b}^{\prime})\cdot {\bf \sigma}\rangle
+\langle {\bf a}^{\prime}\cdot{\bf \sigma}\otimes ({\bf b}-{\bf b}^{\prime})\cdot{\bf \sigma}
\rangle\nonumber\\
&=&\int P(\lambda)d\lambda \{a_{\psi}(\theta,\lambda)[b_{\psi}(\varphi,\lambda)+b_{\psi}(\varphi^{\prime},\lambda)]\nonumber\\
&&+a_{\psi}(\theta^{\prime}, \lambda)[b_{\psi}(\varphi,\lambda)
-b_{\psi}(\varphi^{\prime},\lambda)]\}
\end{eqnarray}
using the simultaneous dispersion-free representations for all the non-commuting operators. By noting the relation
\begin{eqnarray}
a_{\psi}(\theta,\lambda)[b_{\psi}(\varphi,\lambda)+b_{\psi}(\varphi^{\prime},\lambda)]
+a_{\psi}(\theta^{\prime}, \lambda)[b_{\psi}(\varphi,\lambda)
-b_{\psi}(\varphi^{\prime},\lambda)]=\pm 2, \nonumber
\end{eqnarray}
we conclude  for {\em any} $P(\lambda)$
\begin{eqnarray}
|\langle B\rangle_{\psi} |\leq 2.
\end{eqnarray}
Note that this inequality is essentially the same as (5.2).

The hidden-variables model of Bell and
CHSH thus predicts  $|\langle B\rangle_{\psi} |\leq 2\sqrt{2}
$ or $|\langle B\rangle_{\psi} |\leq 2$, for the identical quantum operator $B$ depending on the two different ways of evaluation. 
Physical processes described by $\langle |{\bf b}+{\bf b}^{\prime}|[{\bf a}\cdot {\bf \sigma}\otimes\tilde{{\bf b}}\cdot{\bf \sigma}]\rangle_{\psi}$ and $\langle {\bf a}\cdot {\bf \sigma}\otimes {\bf b}\cdot{\bf \sigma}\rangle_{\psi}+\langle {\bf a}\cdot {\bf \sigma}\otimes {\bf b}^{\prime}\cdot{\bf \sigma}\rangle_{\psi}$ are quite different, but both of them are measurable and quantum mechanics tells that these two should always agree.  

The origin of the above disagreement of the two different ways of evaluation is traced to the failure of the quantum mechanical linearity condition in the hidden-variables model of Bell and CHSH, namely,
\begin{eqnarray}
\langle {\bf a}\cdot {\bf \sigma}\otimes ({\bf b}\pm {\bf b}^{\prime})\cdot {\bf \sigma}\rangle
=\langle {\bf a}\cdot {\bf \sigma}\otimes {\bf b}\cdot {\bf \sigma}\rangle
 \pm \langle {\bf a}\cdot {\bf \sigma}\otimes {\bf b}^{\prime}\cdot {\bf \sigma}\rangle
\end{eqnarray}
for  non-collinear ${\bf b}$ and ${\bf b}^{\prime}$ does not hold in general~\cite{fujikawa2}.
We note that 
linearity is a {\em local property} of quantum mechanics in contrast to entanglement.
\\

One may thus tentatively conclude either\\
\noindent (i) the local hidden-variables model of Bell and CHSH contradicts quantum mechanics due to the failure of linearity  without referring to long-ranged EPR entanglement, or\\
\noindent (ii) one needs to examine the precise consequences of the linearity condition  which renders the conventional CHSH inequality  $|\langle B\rangle_{\psi} |\leq 2$ as the unique prediction of the model.

\subsection{Linearity condition and separability}
We now examine the consequences of linearity condition in detail.
The linearity condition for {\em non-collinear} ${\bf b}$ and ${\bf b}^{\prime}$, which we consider in the following, implies
\begin{eqnarray}
\langle {\bf 1}\otimes ({\bf b}+{\bf b}^{\prime})\cdot {\bf \sigma}\rangle
&=&\int |{\bf b}+{\bf b}^{\prime}|\tilde{b}_{\psi}(\phi,\lambda)P(\lambda)d\lambda\nonumber\\
&=&\int b_{\psi}(\varphi,\lambda)P(\lambda)d\lambda+\int b_{\psi}(\varphi^{\prime},\lambda)P(\lambda)d\lambda
\end{eqnarray}
and 
\begin{eqnarray}
\langle {\bf a}\cdot {\bf \sigma}\otimes ({\bf b}+{\bf b}^{\prime})\cdot {\bf \sigma}\rangle
&=&\int a_{\psi}(\theta,\lambda)|{\bf b}+{\bf b}^{\prime}|\tilde{b}_{\psi}(\phi,\lambda)P(\lambda)d\lambda\nonumber\\
&=&\int a_{\psi}(\theta,\lambda)[b_{\psi}(\varphi,\lambda)+ b_{\psi}(\varphi^{\prime},\lambda)]P(\lambda)d\lambda.
\end{eqnarray}
On the other hand, the well-known von Neumann's no-go argument~\cite{neumann} shows that the expressions local in the $\lambda$ space are quite different, namely,
\begin{eqnarray}
|{\bf b}+{\bf b}^{\prime}|\tilde{b}_{\psi}(\phi,\lambda)\neq b_{\psi}(\varphi,\lambda)+ b_{\psi}(\varphi^{\prime},\lambda)
\end{eqnarray}
which is a general statement on the dispersion-free representations of two non-commuting operators at {\em any point} in hidden-variables space $\lambda$. Note that the only allowed integer for the left-hand side with non-collinear ${\bf b}\ {\rm and}\ {\bf b}^{\prime}$ is $\pm 1$ while the right-hand side is $\pm2\ {\rm or}\ 0$. 
\\

To make the analysis of linearity condition transparent,  we employ the simplest parameterization of the hidden variables and dichotomic variables which was suggested in the original paper of Bell~\cite{bell2},
\begin{eqnarray}
&&P(\lambda)d\lambda=P(\lambda_{1},\lambda_{2})d\lambda_{1}d\lambda_{2},\nonumber\\
&&a_{\psi}(\theta,\lambda)= a_{\psi}(\theta,\lambda_{1}),\ \ \
b_{\psi}(\varphi,\lambda)=  b_{\psi}(\varphi,\lambda_{2}),\nonumber\\
&&\tilde{b}_{\psi}(\phi,\lambda)=  \tilde{b}_{\psi}(\phi,\lambda_{2}),\ \ \
\tilde{b}^{\prime}_{\psi}(\phi^{\prime},\lambda)= \tilde{b}^{\prime}_{\psi}(\phi^{\prime},\lambda_{2}). 
\end{eqnarray}
Namely, the $a$-system is parameterized by the hidden variables $\lambda_{1}$ and the $b$-system is parameterized by the hidden variables $\lambda_{2}$;  each system is described by its own independent local parameters. 
We also define the "projection operator" by 
\begin{eqnarray}
A_{\psi}(\theta,\lambda_{1})=\frac{1}{2}[1 + a_{\psi}(\theta,\lambda_{1})]
\end{eqnarray}
which assumes $1$ or $0$. 

Then the conditions of the linearity (5.20)-(5.22) are summarized by 
\begin{eqnarray}
&&\int [|{\bf b}+{\bf b}^{\prime}|\tilde{b}_{\psi}(\phi,\lambda_{2})
-b_{\psi}(\varphi,\lambda_{2})-b_{\psi}(\varphi^{\prime},\lambda_{2})]
P(\lambda_{1},\lambda_{2})d\lambda_{1}d\lambda_{2}=0,
\end{eqnarray}
\begin{eqnarray}
\int A_{\psi}(\theta,\lambda_{1})[|{\bf b}+{\bf b}^{\prime}|\tilde{b}_{\psi}(\phi,\lambda_{2})
-b_{\psi}(\varphi,\lambda_{2})-b_{\psi}(\varphi^{\prime},\lambda_{2})]
P(\lambda_{1},\lambda_{2})d\lambda_{1}d\lambda_{2}=0,
\end{eqnarray}
and 
\begin{eqnarray}
&&|{\bf b}+{\bf b}^{\prime}|\tilde{b}_{\psi}(\phi,\lambda_{2})
-b_{\psi}(\varphi,\lambda_{2})-b_{\psi}(\varphi^{\prime},\lambda_{2})\neq 0,
\end{eqnarray}
for any pure state $\psi$ and for any  free parameters  $\theta$, $\varphi$ and $\varphi^{\prime}$, and in the case of (5.27) for any $\lambda_{2}$.

We determine the possible structure of the weight function $P(\lambda_{1},\lambda_{2})$ from the conditions (5.25)-(5.27). From (5.25) and (5.26), we see that 
\begin{eqnarray}
&&P(\Lambda_{1}; \lambda_{2})=\int_{\Lambda_{1}}d\lambda_{1}P(\lambda_{1},\lambda_{2}), \nonumber\\
&&P(\psi, \theta; \lambda_{2})=\int_{\Lambda_{1}}d\lambda_{1}A_{\psi}(\theta,\lambda_{1})P(\lambda_{1},\lambda_{2}), \nonumber\\
&&\bar{P}(\psi, \theta; \lambda_{2})=\int_{\Lambda_{1}}d\lambda_{1}[1-A_{\psi}(\theta,\lambda_{1})]P(\lambda_{1},\lambda_{2}), 
\end{eqnarray}
define the weight factors of consistent (i.e., satisfying linearity) $d=2$ hidden-variables models of the $b$-system, where $\Lambda_{1}$ is the entire space of the variables $\lambda_{1}$. The non-negative weight $P(\psi, \theta; \lambda_{2})$ receives the contribution from the domain with $A_{\psi}(\theta,\lambda_{1})=1$, and the non-negative  weight $\bar{P}(\psi, \theta; \lambda_{2})$ from its complement. For non-trivial hidden-variables models, we have $P(\psi, \theta; \lambda_{2})\neq P(\Lambda_{1}; \lambda_{2})$ in general. In particular, the first relation of (5.28) shows that $P(\Lambda_{1}; \lambda_{2})$ defines $d=2$ {\em non-contextual} hidden-variables models.
Due to the symmetry between $a$-system and $b$-system, we can make a similar statement on $\lambda_{1}$ dependence.

If one assumes that the weights for the $d=2$ non-contextual hidden-variables models are {\em uniquely} specified by the chosen dichotomic representations of $a_{\psi_{1}}(\theta,\lambda_{1})$ and $b_{\psi_{2}}(\varphi,\lambda_{2})$, respectively, which is the case of the known construction of hidden-variables models in $d=2$~\cite{bell1, kochen}, one concludes from (5.28) a factored form of two systems
\begin{eqnarray}
P(\lambda_{1},\lambda_{2})=P_{1}(\lambda_{1})P_{2}(\lambda_{2})
\end{eqnarray}
where $P_{1}(\lambda_{1})$ stands for the weight of the $a$-system and 
$P_{2}(\lambda_{2})$ stands for the weight of the $b$-system in the sense of consistent non-contextual hidden-variables models in $d=2$.  
For this choice in (5.29), $P(\psi, \theta; \lambda_{2})$ and $P(\Lambda_{1}; \lambda_{2})$ are equivalent as the weight for the $b$-system, and similarly for the $a$-system. The analysis of more general parameterization than (5.23) is found in~\cite{fujikawa2}.

We thus arrive at the conclusion that the formula of Bell and CHSH is now written as 
\begin{eqnarray}
\langle {\bf a}\cdot {\bf \sigma}\otimes {\bf b}\cdot {\bf \sigma}\rangle_{\psi}
=\int_{\Lambda_{1}} P_{1}(\lambda_{1})a_{\psi}(\theta,\lambda_{1})d\lambda_{1}
\int_{\Lambda_{2}}P_{2}(\lambda_{2})b_{\psi}(\varphi,\lambda_{2})d\lambda_{2}
\end{eqnarray}
namely, the formula is {\em valid only for the pure separable state}, 
\begin{eqnarray}
\rho=|\psi_{1}\rangle\langle\psi_{1}|\otimes|\psi_{2}\rangle\langle\psi_{2}|,
\end{eqnarray}
and it is not applicable to entangled states~\cite{fujikawa2}.  

We emphasize that a factored product of two non-contextual $d=2$ hidden variables models~\cite{bell1,kochen} does not contradict Gleason's theorem which prohibits a non-contextual $d=4$ hidden-variables model.

Also the conclusion in the original paper of Bell~\cite{bell2}, which states that an entangled singlet state is inconsistent with his hidden-variables model, is natural since our analysis shows that the hidden-variables model in~\cite{bell2} is applicable only to separable quantum mechanical states. 

\subsection{Implications of conventional CHSH inequality}

The CHSH inequality follows from the fact that the local hidden-variables model of Bell and CHSH in (5.8) is valid only for pure separable states. 
For pure separable quantum states, it is readily confirmed that the ordinary CHSH inequality 
\begin{eqnarray}
|\langle B\rangle| \leq 2
\end{eqnarray}
naturally holds, as we have explained in (5.7). Our conclusion is thus perfectly consistent with the analysis of Werner~\cite{werner}, who shows that CHSH inequality is a necessary and sufficient separability condition of pure quantum mechanical states.
\\

As an 
interesting practical application of CHSH inequality, we mention the application to {\em quantum cryptography} by Ekert~\cite{ekert}, which is based on the mixed separable quantum states 
\begin{eqnarray}
\rho=\int d{\bf n}_{a}d{\bf n}_{b}w({\bf n}_{a},{\bf n}_{b})\rho({\bf n}_{a})\otimes\rho({\bf n}_{b}).
\end{eqnarray} 
This $\rho$ satisfies  the relation  
\begin{eqnarray}
-2\leq{\rm Tr}[\rho B]\leq2.
\end{eqnarray}
It is crucial that {\em no dispersion-free representations} appear in this consideration. Classical vector quantities ${\bf n}_{a}$ and ${\bf n}_{b}$ have {\em no direct connection} with hidden variables. This application is based on the fact that any separable quantum mechanical states satisfy CHSH inequality and it has nothing to do with the hidden-variables model.

Conventional CHSH inequality  $|\langle B\rangle |\leq 2$ does not provide a test of the local non-contextual hidden-variables model in $d=4$.  The non-contextual hidden-variables model in $d=4$ simply does not exist as Gleason's theorem implies.

\section{Conclusion}
We have critically reassessed hidden-variables models in connection with quantum discord and CHSH inequality. We have shown that the original hidden-variables model of Bell in $d=2$ fails to describe the conditional measurement consistently. Thus it fails to describe the criterion of  vanishing quantum discord in a classical deterministic manner. This is consistent with the fact that quantum discord is a quantum mechanical effect.  
We have also shown that the local non-contextual hidden-variables model of Bell and CHSH in $d=4$, which was introduced to show that the local hidden-variables model cannot describe entanglement, in fact does not define a non-trivial non-contextual model in $d=4$. In particular, we have shown that the conventional CHSH inequality $|\langle B\rangle| \leq 2$ is simply a manifestation of separable quantum mechanical states and does not constitute a test of $d=4$ non-contextual hidden-variables model, in agreement with Gleason's theorem.

As for hidden-variables models, we conclude that {\em no viable models of local non-contextual hidden-variables exist in 
any dimensions of Hilbert space.}


\begin{thebibliography}{99}

\bibitem{zurek}
H. Ollivier and W. H.  Zurek, Phys. Rev. Lett. {\bf 88}, 017901 (2002) .
\bibitem{vedral}
L. Henderson and V. Vedral, J. Phys. A{\bf 34}, 6899 (2001).
\bibitem{bell1}
J. S. Bell, Rev. Mod. Phys. {\bf 38}, 447 (1966).
\bibitem{kochen}
S. Kochen and E. P. Specker, J. Math. Mech. {\bf 17}, 59 (1967).
\bibitem{beltrametti}
E. G. Beltrametti and G. Gassinelli, {\em The Logic of Quantum 
Mechanics}, (Addison-Wesley Pub., 1981).
\bibitem{peres}
A. Peres, {\em Quantum Theory: Concepts and Methods }, (Kluwer Academic Pub., 1995).
\bibitem{fujikawa1}
K. Fujikawa, Phys. Rev. A85 (2012) 012114.
\bibitem{bell2}
J. S. Bell, Physics {\bf 1}, 195 (1965).
\bibitem{chsh}
J. F. Clauser, M. A. Horne, A. Shimony and R. A. Holt, Phys. Rev. Lett. 
{\bf 23},  888 (1969).
\bibitem{cs}
The early history of Bell and CHSH inequalities is well reviewed
in, J.F. Clauser and A. Shimony, Rep. Prog. Phys. {\bf 41}, 1881 (1978).
\bibitem{cirelson}
B.S. Cirel'son, Lett. Math. Phys. {\bf 4} (1980) 93.
\bibitem{fujikawa2}
K. Fujikawa, Prog. Theor. Phys. 127 (2012), 975.
\bibitem{gleason}
A. M. Gleason, J. Math. Mech. {\bf 6}, 885 (1957).
\bibitem{aspect}
A. Aspect, J. Dalibard and G. Roger, Phys. Rev. Lett. {\bf 49},
1804 (1982).
\bibitem{ekert}
A.K. Ekert, Phys. Rev. Lett. {\bf 67}, 661 (1991).
\bibitem{umegaki}
H. Umegaki, Tohoku Math. J. {\bf 6}, 177 (1954).
\bibitem{davies}
E. B. Davies and J. T. Lewis, Comm. Math. Phys. {\bf 17}, 239 (1970).  
\bibitem{fujikawa3}
K. Fujikawa, Prog. Theor. Phys. 127 (2012), 989.
%\bibitem{wehrl}
%A. Wehrl, Rev. Mod. Phys. {\bf 50}, 221 (1978).
\bibitem{datta1}
A. Datta, {\em Ph. D. Thesis}, University of New Mexico (2008), arXiv:0807.4490v1.
\bibitem{dakic}
B. Dakic, V. Vedral and C. Brukner, Phys. Rev. Lett. {\bf 105}, 190502 (2010).
\bibitem{datta2}
A. Datta, arXiv:1003.5256v1[quant-ph].
\bibitem{neumann}
J. von Neumann, {\em Mathematical Foundations of Quantum Mechanics}
(Princeton Univ. Press, 1955).
\bibitem{werner}
R.F. Werner, Phys. Rev. A{\bf 40}, 4277 (1989).
\end{thebibliography}
\end{document}